\documentclass[
reprint,
superscriptaddress,
nofootinbib,
 amsmath,amssymb,
 aps,
 prl,
]{revtex4-2}

\usepackage[normalem]{ulem}
\usepackage{xcolor}
\usepackage{upgreek}
\usepackage[T1]{fontenc}
\usepackage{graphicx}
\usepackage{dcolumn}
\usepackage{bm}
\usepackage{hyperref}
\usepackage{subfigure}
\hypersetup{
    colorlinks=true,
    linkcolor=magenta,
    filecolor=red,
    urlcolor=blue,
    pdftitle={Precision measurement of coherent elastic neutrino-nucleus scattering at a nuclear reactor by CONUS+},
}
\usepackage[all]{hypcap}
\usepackage[inter-unit-product =\cdot]{siunitx}
\usepackage{gensymb}
\usepackage[mathlines]{lineno}

\newcommand{\CEvNS}{CE$\nu$NS}
\newcommand{\conusplus}{CONUS+}

\makeatletter
\let\frontmatter@footnote@produce\frontmatter@footnote@produce@footnote
\def\@fnsymbol#1{\ensuremath{\ifcase#1 \or \dagger \or \dagger\dagger \or \ddagger \or \ddagger\ddagger \or * \or \mathsection \or \mathsection\mathsection \or ** \or \parallel \or \mathparagraph \else\@ctrerr\fi}}
\makeatother

\begin{document}

\title{Precision measurement of coherent elastic neutrino-nucleus scattering at a nuclear reactor by \conusplus}

\newcommand{\MPIK}{\affiliation{Max-Planck-Institut f\"ur Kernphysik, Saupfercheckweg 1, 69117 Heidelberg, Germany}}
\newcommand{\KIT}{\affiliation{Institut f\"ur Astroteilchenphysik, Karlsruher Institut f\"ur Technologie (KIT), Hermann-von-Helmholtz-Platz 1, 76344 Eggenstein-Leopoldshafen, Germany}}
\newcommand{\KKL}{\affiliation{Kernkraftwerk Leibstadt AG, 5325 Leibstadt, Switzerland}}
\newcommand{\MBI}{\altaffiliation[Present address: ]{Marietta-Blau-Institut f\"ur Teilchenphysik der \"OAW, Dominikanerbastei 16, 1010 Wien, Austria}}
\newcommand{\Mirion}{\altaffiliation[Present address: ]{Mirion Technologies GmbH, Stahlstra\ss e 42-44, 65428 R\"usselsheim, Germany}}
\newcommand{\PSI}{\altaffiliation[Present address: ]{Paul Scherrer Institut, Forschungsstrasse 111, 5232 Villigen, Switzerland}}

\newcommand{\ESGmail}{\altaffiliation[Contact author: ]{\protect\href{mailto:esanchez@mpi-hd.mpg.de}{esanchez@mpi-hd.mpg.de}}}

\author{N.~Ackermann}\MPIK
\author{H.~Bonet}\MPIK
\author{C.~Buck}\MPIK
\author{J.~Hakenm\"uller}\MBI\MPIK
\author{G.~Heusser}\MPIK
\author{M.~Lindner}\MPIK
\author{W.~Maneschg}\Mirion\MPIK
\author{M.~Meier}\MPIK
\author{S.~Mertens}\MPIK
\author{D.~Piani}\MPIK
\author{M.~Rank}\KKL
\author{T.~Rink}\MPIK\KIT
\author{E.~S\'{a}nchez Garc\'{i}a}\ESGmail\MPIK
\author{Y.~Shi}\MPIK
\author{L.~Sch\"onfeld}\MPIK
\author{I.~Stalder}\KKL
\author{H.~Strecker}\MPIK
\author{J.~Wendel}\MPIK
\author{J.~Woenckhaus}\PSI\KKL

\collaboration{CONUS Collaboration}
\email[Contact author: ]{conus.eb@mpi-hd.mpg.de}

\date{\today}

\begin{abstract}
We report the most precise measurement to date of coherent elastic neutrino-nucleus scattering (\CEvNS) from reactor antineutrinos, using the \conusplus~germanium detectors at the Leibstadt nuclear power plant. With an exposure of 1296~kg$\cdot$days reactor-on, a blind likelihood analysis yields a signal with high statistical significance, corresponding to $1565 \pm 155$ \CEvNS~events, in agreement with the Standard Model expectation of $1701 \pm 198$ events within 0.5~$\sigma$. The flux-averaged cross section is $(0.57 \pm 0.09)\times10^{-40}$\,cm$^2$, and the weak mixing angle at a momentum transfer of about 10~MeV is $\sin^2\theta_W = 0.226 \pm 0.022$.
\end{abstract}

\maketitle

Coherent elastic neutrino-nucleus scattering (\CEvNS) is a Standard Model (SM) process predicted in 1974~\cite{Freedman:1973yd, Kopeliovich:1974mv}. Its cross section, enhanced by coherent scattering off the whole nucleus, enables compact, kilogram-scale detectors. The first \CEvNS~experimental detection was achieved by the COHERENT Collaboration using neutrinos of up to about 50~MeV from a spallation neutron source~\cite{Coherent:2017,COHERENT:2020iec,COHERENT:2026yje}. Evidence of \CEvNS~from $^{8}$B solar neutrinos, with energies up to 15~MeV, has also been reported~\cite{XENON:2026smt,PandaX:2024muv, LZ:2025igz}, marking the onset of the neutrino fog for future dark matter searches. Nuclear reactors are a complementary \CEvNS~source, producing antineutrinos of lower energy~$E_{\nu}\lesssim10$\,MeV. There is currently a broad worldwide effort to detect \CEvNS~at reactors using different targets~\cite{CONNIE:2024pwt, MINER:2025cevns, Colaresi:2022obx, NEON:2022hbk, NUCLEUS:2026pnv, nGeN:2025hsd, Ackermann:2024kxo, Akimov:2024lnl, Bailly-Salins:2026inm, Dai:2025kai, Xie:2025frh, TEXONO:2026eyr},  with the first detection of reactor \CEvNS~reported by \conusplus~at a significance of 3.7~$\sigma$~\cite{Ackermann:2025obx}, in its first measurement campaign (run 1). The uncertainty on the extracted signal was dominated by statistics, while the uncertainty on the prediction was dominated by the calibration of the energy scale near the detector threshold~\cite{Ackermann:2025obx}.

\conusplus~is installed at the Leibstadt nuclear power plant (Kernkraftwerk Leibstadt; KKL), Switzerland, at a distance of 20.7~m from the 3.6~GW$_\mathrm{th}$ reactor core, and has been in operation since November 2023~\cite{CONUS:2024lnu}. The site provides an average overburden of 7.4~m of water equivalent~\cite{CONUS:2024vyx}. Since the end of run 1, the detector array has been upgraded, increasing the active target mass by about 50\% and hence the expected signal rate. This Letter reports the \conusplus~run 2 result, based on an about four times larger exposure of 1296~kg$\cdot$days reactor-on, which yields the most precise measurement of reactor \CEvNS~to date and an updated determination of the weak mixing angle.

\textit{Experimental setup.}---In run 1, four 1~kg p-type point-contact high-purity germanium detectors were operated. For run 2, three were replaced by 2.42~kg diodes, of which two, C6 and C7, were operational after installation, with active masses of $(2.33^{+0.02}_{-0.03})$~kg each. Including the original detector C3, with $(0.94 \pm 0.01)$~kg, the total active mass is $(5.60^{+0.03}_{-0.04})$~kg. The detectors are housed inside a compact shield combining layered lead and borated polyethylene, an active muon veto, and continuous radon-free air flushing, which together suppress the background by four orders of magnitude~\cite{Bonet:2021wjw, Ackermann:2026vkq}. Signals from the transistor-reset preamplifiers are continuously digitized, and the energy is reconstructed online with a trapezoidal filter, as described in~\cite{CONUS:2024lnu}.

Reactor antineutrinos produce germanium nuclear recoils of at most a few keV, of which only about 15--20\% is converted into ionization due to quenching~\cite{Bonhomme:2022lcz}, expressed in electron-equivalent units (eV$_{ee}$). The \CEvNS~signal is therefore concentrated just above the energy threshold of each detector (Table~\ref{tab:results}), within a region of interest (ROI) from the threshold up to 800~eV$_{ee}$, making the energy scale and the quenching factor critical for the signal prediction. Despite their larger size, the new detectors have a noise level comparable to the original ones, with a pulser full width at half maximum (FWHM) of about 50~eV$_{ee}$ and full trigger efficiency above 180~eV$_{ee}$. The energy scale is calibrated with atomic de-excitation lines of the $^{71}$Ge produced in situ~\cite{SanchezGarcia:2026nvr}, with an uncertainty at threshold of 3~eV$_{ee}$, stable over the full run. Below the lowest calibration line at 1.3~keV$_{ee}$, pulser injections show a non-linearity of the energy scale of less than 4~eV$_{ee}$.

\textit{Data acquisition.}---Run 2 data were collected from February 2025 to July 2026, excluding periods with instabilities in the radon-flushing system or the electrical grounding. The electronic noise rate remained stable within 10\% and the noise FWHM within 2~eV$_{ee}$. The background rate in the ROI, evaluated after unblinding, was likewise stable, with a slight reduction during reactor-off periods due to the additional overburden of the reactor drywell lid, which is accounted for in the background model~\cite{Ackermann:2025obx, Ackermann:2026vkq}. After event selection and dead-time correction, the exposure is 1296~kg$\cdot$days reactor-on and 86~kg$\cdot$days reactor-off~\cite{CONUS:2026run2}.

\textit{Background model.}---The run 2 background consists of seven components: muon- and neutron-induced events, cosmogenic activation of the crystals and of the copper cryostats, radon, inert gases released by the reactor, and $^{210}$Pb in the shield. It is simulated within the Geant4-based framework MaGe~\cite{Boswell:2010mr} and fitted simultaneously to the measured spectra of each detector above 0.4~keV$_{ee}$, where the expected \CEvNS~signal is negligible. The muon- and neutron-induced contributions are based on a dedicated characterization of the experimental site~\cite{CONUS:2024vyx}. Charge-collection effects and Lindhard quenching are applied event-by-event in a post-processing step. For the new detectors, the manufacturer specifies a partial charge collection layer (PCCL) thickness of $(0.6^{+0.2}_{-0.1})$\,mm. With this constraint, a comparison of cosmic-muon simulations with run 2 data taken without the muon veto cut yields dead-layer (DL) and transition-layer (TL) thicknesses of about 0.25 and 0.35~mm~\cite{CONUS:2026run2}. For C3, previous $^{241}$Am attenuation measurements determined 0.48~mm and 0.16~mm, respectively~\cite{Bonet:2020ntx}. A combined fit to the background-only region from 0.4~keV$_{ee}$ to 28~keV$_{ee}$, together with the range up to 400~keV$_{ee}$ constraining the radon-induced $\gamma$ lines and metastable-germanium peaks, provides priors for the normalizations of the seven background components entering the likelihood. The background in the ROI is dominated by cosmic-ray neutrons, while above 100~keV$_{ee}$ muon-induced events dominate, followed by $^{210}$Pb and radon. The new, larger detectors achieve a background level per unit mass reduced by approximately a factor of two relative to the original 1~kg diodes, due to more radiopure cryostat materials and the more favorable mass-to-surface ratio of the larger crystals~\cite{CONUS:2026run2}.

\textit{Signal prediction.}---The expected \CEvNS~signal for run 2 is calculated using the same framework as in run 1~\cite{Ackermann:2025obx}, updated with the reactor thermal power and fission fractions of the run 2 period. The nuclear recoil spectrum is converted into ionization energy using the Lindhard model with $k=0.162\pm0.004$~\cite{Bonhomme:2022lcz}. Integrated over the ROI, the predicted signal for the combined run 2 exposure is $N_{\mathrm{SM}} = 1701 \pm 198$ events, with the prediction for each detector given in Table~\ref{tab:results}. The total relative uncertainty on the prediction is reduced from 16.4\% in run 1 to 11.6\%. This improvement comes mainly from the more precise energy-scale calibration with a reduced uncertainty at threshold of 8.0\% instead of 14.1\%, now comparable to that of the quenching factor (7.5\%).

\textit{Data analysis.}---The \CEvNS~signal is extracted through a binned Poisson likelihood fit performed simultaneously on the reactor-on and reactor-off spectra of the three detectors. The parameter of interest is the number of \CEvNS~counts $N_s$ in the ROI, summed over all detectors. Gaussian pull terms constrain nuisance parameters to their central values and uncertainties, including trigger efficiency, energy calibration, active mass, and normalizations of the background components obtained from the background model, which are shared between the reactor-on and reactor-off periods. The statistical uncertainty of the simulated background spectra is included following~\cite{Barlow:1993dm}. Uncertainties on the signal prediction, such as the reactor flux, do not enter the fit and are only included when comparing $N_s$ to the SM expectation. Two independent likelihood implementations, differing in the treatment of the energy resolution and of some systematic uncertainties, were developed. Likelihood A is used as default. Likelihood B additionally includes the uncertainties on the non-linearity and on the shape of the reactor antineutrino spectrum. These are added in quadrature to the result of Likelihood A, and the difference between both implementations is included as an additional systematic uncertainty. Full details are given in~\cite{CONUS:2026run2}.

We followed a blind-analysis strategy, in which the measured spectra below 400~eV$_{ee}$ were hidden. The likelihood modifications introduced for run 2 were first validated on the published run 1 dataset and then tested on Asimov datasets. The detectors were then unblinded one by one, comparing the results of Likelihood A and B at each step. Five modifications were made during and after the unblinding, each motivated by checks independent of the signal, as described in~\cite{CONUS:2026run2}.

\textit{Sensitivity.}---Prior to unblinding, the expected sensitivity was evaluated on Asimov datasets with the run 2 inputs. At the full run 2 exposure, the expected value of $N_s/\sigma(N_s)$ for the SM \CEvNS~signal is 10.5, with an expected relative uncertainty on $N_s$ of about 9.5\%, of which 7.3\% is statistical and 6.1\% systematic. Including the uncertainty on the signal prediction, the expected uncertainty on the measured cross section relative to the SM is about 15\%. 

\begin{figure}[t]
\includegraphics[width=\columnwidth]{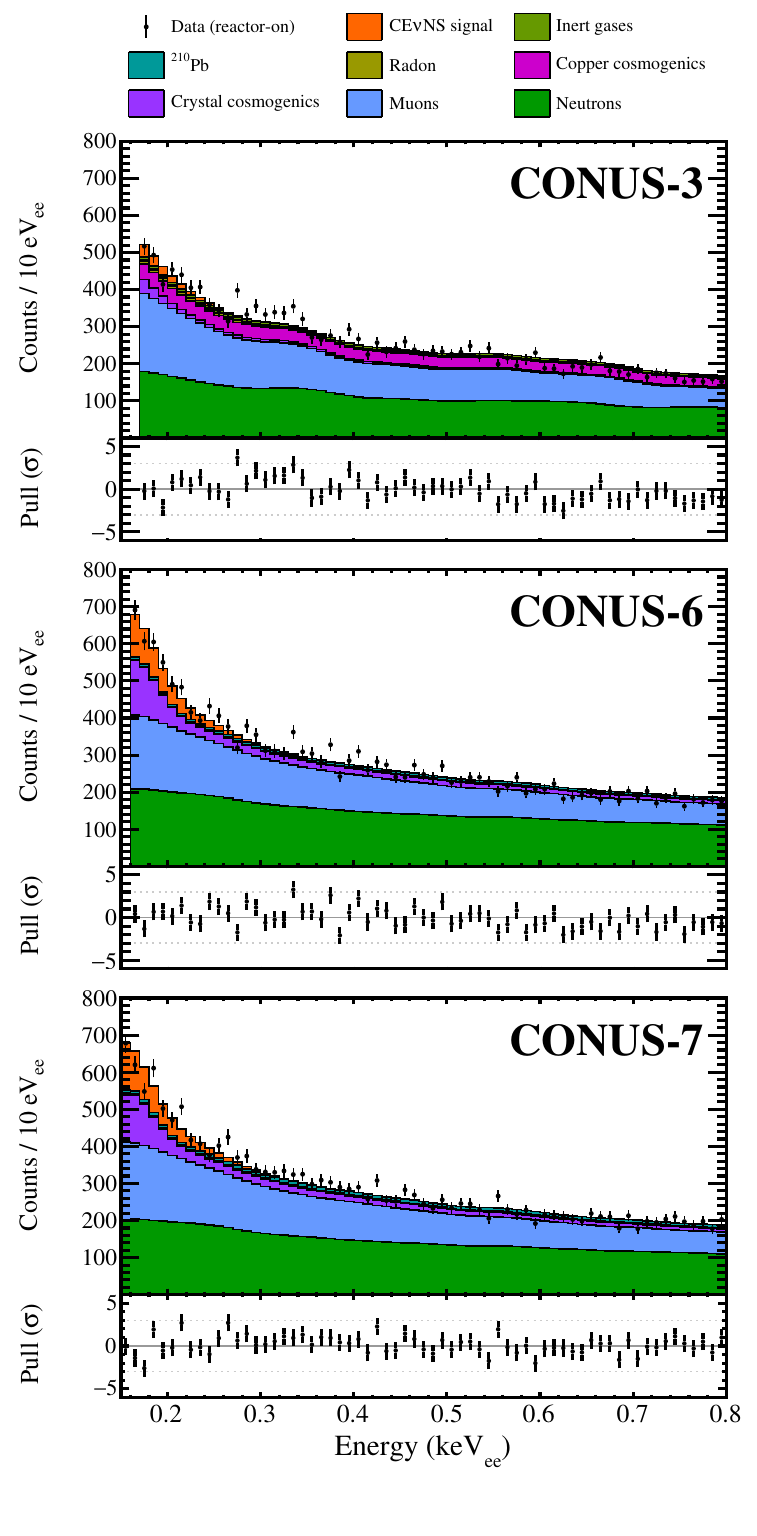}
\caption{Reactor-on spectra of C3 (top), C6 (middle), and C7 (bottom) in the ROI, together with the background components and the \CEvNS~signal (orange) from the combined fit of the three detectors. The lower panels show the pulls between data and model. The reactor-off spectra and the individual detector fits are shown in~\cite{CONUS:2026run2}.}
\label{fig:result}
\end{figure}

\textit{Results.}---Figure~\ref{fig:result} shows the reactor-on spectra of the three detectors in the ROI, together with the background components and the \CEvNS~signal from the combined fit of the three detectors. The combined fit yields $N_s = 1565 \pm 117\,\mathrm{(stat)} \pm 102\,\mathrm{(syst)}$ \CEvNS~events, corresponding to a relative precision of 9.9\% ($N_s/\sigma(N_s) = 10.1$), in agreement with the expected sensitivity, and Likelihood B yields a consistent result of $N_s = 1575 \pm 150$. The individual detector fits, summarized in Table~\ref{tab:results}, give $N_s/\sigma(N_s)$ of 6.5 (C6), 8.0 (C7), and 2.4 (C3), and agree with the SM expectation within 1.1 standard deviations.

The combined fit has $\chi^2$/ndf = 1.39 for 381 effective degrees of freedom ($p = 3\times10^{-6}$, from pseudo-experiments). The deviations arise mainly in C3, which contributes about 11\% of the expected signal, and in the reactor-off data of C6, while C7, with the largest expected signal (53\%), is well described. The impact on the result was assessed with several checks~\cite{CONUS:2026run2}. A fit of the reactor-on minus reactor-off spectra, which does not rely on the simulated background shape, yields $N_s = 1318 \pm 403$, corresponding to a significance of 3.3~$\sigma$ and consistent with the combined fit within the uncertainties. A fit of the reactor-off data with a free signal component yields $N_s = 23 \pm 30$, consistent with zero, whereas a reactor-independent signal would give about 103 counts. The ratio $N_s/N_{\mathrm{SM}}$ remains stable when raising the energy thresholds by 10 and 20~eV$_{ee}$ (0.92 and 0.97) and changes by less than 9\% for extended fit ranges. Finally, the statistical variance of all bins was scaled by a common factor until $p > 0.1$, reached at a factor of 1.3, which increases the relative uncertainty on $N_s$ to 10.9\% ($N_s/\sigma(N_s) = 9.2$).

\begin{figure}[t]
\includegraphics[width=\columnwidth]{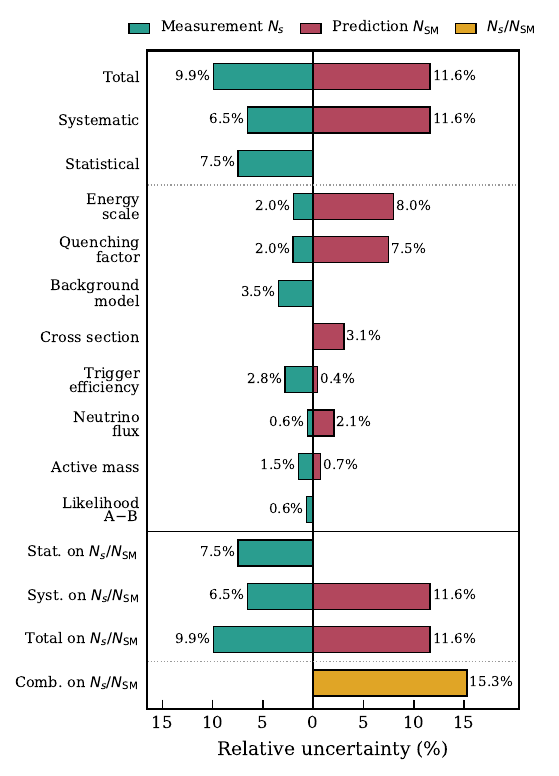}
\caption{Relative uncertainty on $N_s$ from the combined fit (left) and on $N_{\mathrm{SM}}$ (right), in total and by source. Each contribution to $N_s$ is the quadrature reduction of the total uncertainty when the corresponding nuisance parameters are fixed. The energy scale includes the non-linearity, and the neutrino flux includes the spectrum shape. Due to correlations between nuisance parameters, the contributions to $N_s$ do not add in quadrature to the total systematic uncertainty. The lower rows show the uncertainties entering $N_s/N_{\mathrm{SM}}$ and their quadrature combination (Comb.).}
\label{fig:breakdown}
\end{figure}

The total relative uncertainty on $N_s$ of 9.9\% is statistics-dominated (7.5\% statistical, 6.5\% systematic). Figure~\ref{fig:breakdown} shows the breakdown of the systematic contributions to $N_s$, together with those of the signal prediction. The largest systematic contributions to $N_s$ arise from the background model (3.5\%) and the trigger efficiency (2.8\%). The energy scale and the quenching factor affect $N_s$ only mildly, as they enter the fit only through the signal shape, but dominate the uncertainty on the prediction (8.0\% and 7.5\%). Compared with $N_{\mathrm{SM}} = 1701 \pm 198$, the measurement yields $N_s/N_{\mathrm{SM}} = 0.92 \pm 0.14$, combining both uncertainties in quadrature and neglecting their correlations. Unlike $N_s$, the ratio is systematics-dominated (13.3\% vs.\ 7.5\% statistical), arising mostly from the uncertainty on the signal prediction, which therefore limits the determination of the weak mixing angle and constraints on physics beyond the SM.

\begin{table}[t]
\caption{Number of \CEvNS~events $N_s$ from the individual detector fits and from the combined fit, compared to the SM prediction $N_{\mathrm{SM}}$, together with the ratio $N_s/\sigma(N_s)$, the reactor-on exposure, and the energy threshold $E_{\mathrm{thr}}$.}
\label{tab:results}
\setlength{\tabcolsep}{3pt}
\renewcommand{\arraystretch}{1.7}
\begin{ruledtabular}
\begin{tabular}{lccccc}
Fit & Exp. & $E_{\mathrm{thr}}$ & $N_s$ & $N_{\mathrm{SM}}$ & $N_s/\sigma(N_s)$ \\[-6pt]
    & (kg$\cdot$d) & (eV$_{ee}$) & & & \\
\hline
C6    & 544  & 160 & $707 \pm 108$  & $604 \pm 73$   & 6.5 \\
C7    & 539  & 140 & $743 \pm 93$   & $903 \pm 111$  & 8.0 \\
C3    & 213  & 170 & $191 \pm 81$   & $195 \pm 23$   & 2.4 \\
\hline
Total & 1296 & --  & $1565 \pm 155$ & $1701 \pm 198$ & 10.1 \\
\end{tabular}
\end{ruledtabular}
\end{table}

\begin{figure}[t]
\includegraphics[width=\columnwidth]{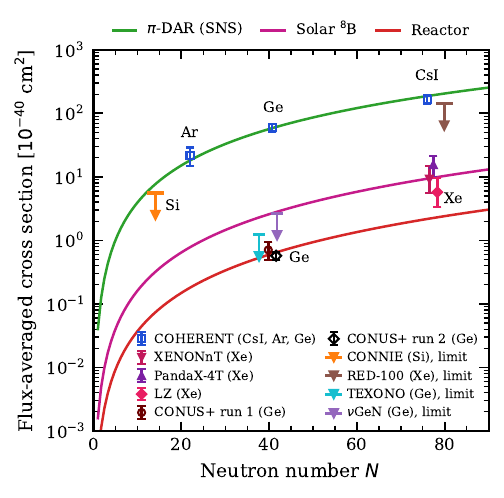}
\caption{Flux-averaged \CEvNS~cross section as a function of the neutron number of the target nucleus, for spallation-source neutrinos, solar $^{8}$B neutrinos, and reactor antineutrinos. The curves show the SM predictions, with the reactor flux-averaged cross section defined for antineutrino energies above 1.8~MeV. The \conusplus~run 1~\cite{Ackermann:2025obx} and run 2 results on germanium are shown together with measurements by COHERENT~\cite{Coherent:2017,COHERENT:2020iec,COHERENT:2026yje} and dark matter experiments~\cite{XENON:2026smt,PandaX:2024muv,LZ:2025igz}, and upper limits from other reactor experiments~\cite{CONNIE:2024pwt,Akimov:2024lnl,TEXONO:2026eyr,nGeN:2025hsd}. Points for germanium are slightly displaced horizontally for visibility.}
\label{fig:xsec}
\end{figure}

\textit{Comparison with other measurements.}---Figure~\ref{fig:xsec} shows the flux-averaged \CEvNS~cross section as a function of the neutron number of the target nucleus, comparing the present result with measurements at spallation sources~\cite{Coherent:2017,COHERENT:2020iec,COHERENT:2026yje}, with solar $^{8}$B neutrinos~\cite{XENON:2026smt,PandaX:2024muv,LZ:2025igz}, and with other reactor experiments~\cite{Ackermann:2025obx,CONNIE:2024pwt,Akimov:2024lnl,TEXONO:2026eyr,nGeN:2025hsd}. The SM predictions are computed with the Helm form factor~\cite{Helm:1956zz}, which is relevant at the higher momentum transfers of spallation-source neutrinos and negligible at reactor energies. The \conusplus~value is obtained from the measured ratio $N_s/N_{\mathrm{SM}}$, scaling the SM flux-averaged cross section, which yields $\langle\sigma\rangle = (0.57 \pm 0.09)\times10^{-40}$\,cm$^2$ for antineutrino energies above 1.8\,MeV. With $1565 \pm 155$ \CEvNS~events, \conusplus~run 2 provides the largest sample of detected \CEvNS~interactions to date. It reaches a precision comparable to that of the recent high-precision COHERENT germanium measurement~\cite{COHERENT:2026yje}, and it is the most precise measurement at reactor antineutrino energies to date. The result is in agreement with the SM expectation and with the previous upper limits on germanium from reactor experiments~\cite{Ackermann:2024kxo,TEXONO:2026eyr,nGeN:2025hsd}. The signal reported by Dresden-II~\cite{Colaresi:2022obx}, interpreted with a quenching factor exceeding the Lindhard model at sub-keV recoil energies~\cite{Collar:2021fcl}, corresponds to a rate significantly higher~\cite{Ackermann:2025obx} than those observed by other germanium experiments at reactors~\cite{Ackermann:2024kxo,Ackermann:2025obx,TEXONO:2026eyr,nGeN:2025hsd}, including the present measurement.

\begin{figure}[t]
\includegraphics[width=\columnwidth]{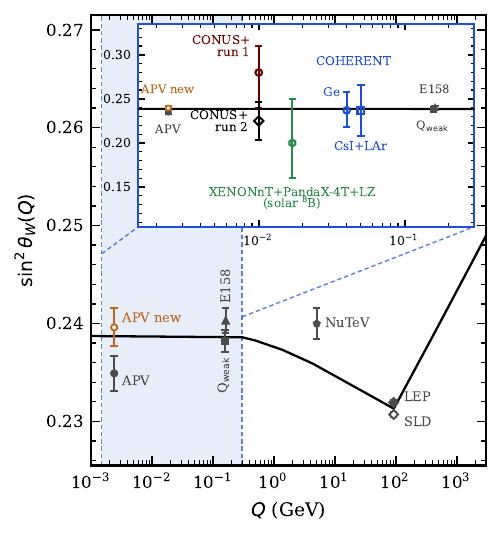}
\caption{Weak mixing angle $\sin^2\theta_W$ as a function of the momentum transfer $Q$. The inset shows the shaded region with an extended vertical scale, including the \conusplus~run 1~\cite{CONUS:2026uhz} and run 2 results together with determinations from atomic parity violation, parity-violating electron scattering, and \CEvNS~at spallation sources and with solar $^{8}$B neutrinos.}
\label{fig:weinberg}
\end{figure}

\textit{Weak mixing angle.}---In the SM, the \CEvNS~cross section is proportional to the square of the weak nuclear charge, $Q_W = N - (1 - 4\sin^2\theta_W)\,Z$, where $N$ and $Z$ are the neutron and proton numbers of the target nucleus. The measurement can therefore be translated into a determination of the weak mixing angle at the low momentum transfer probed by reactor antineutrinos, $Q \approx 10$~MeV, corresponding to antineutrino energies above about 5~MeV at the detector thresholds, as done for run 1~\cite{CONUS:2026uhz}. From the measured ratio $N_s/N_{\mathrm{SM}}$, including the uncertainties on the measurement and on the signal prediction, we obtain $\sin^2\theta_W = 0.226 \pm 0.022$, in agreement within 0.6 standard deviations with the SM expectation of $\sin^2\theta_W = 0.2387$ at low momentum transfer~\cite{ParticleDataGroup:2024cfk}. Figure~\ref{fig:weinberg} compares this result with other determinations, from atomic parity violation~\cite{Wood:1997zq,Guena:2004sq,AtzoriCorona:2024rtv}, parity-violating electron scattering~\cite{E158:2005uwm,Qweak:2018tjf}, neutrino--nucleus deep inelastic scattering~\cite{NuTeV:2001whx}, the $Z$ pole~\cite{ALEPH:2005ab}, and \CEvNS~at spallation sources~\cite{DeRomeri:2022twg,AtzoriCorona:2026wbu} and with solar $^{8}$B neutrinos~\cite{DeRomeri:2026prc}. Among the \CEvNS~determinations, \conusplus~probes the lowest momentum transfer.

\textit{Conclusions and outlook.}---With an exposure of 1296~kg$\cdot$days reactor-on, \conusplus~measures $N_s = 1565 \pm 155$ \CEvNS~events from reactor antineutrinos on germanium, the most precise measurement of reactor \CEvNS~to date. The result is in agreement with the SM expectation, with $N_s/N_{\mathrm{SM}} = 0.92 \pm 0.14$, and consistent with run 1, while the relative uncertainty on the number of signal counts is reduced by almost a factor of three. The measurement also provides a determination of the weak mixing angle, $\sin^2\theta_W = 0.226 \pm 0.022$, at the lowest momentum transfer among \CEvNS~determinations.

Since the measurement remains statistics-dominated, one more year of data would reduce the statistical uncertainty to about 5\%. Pulse-shape discrimination applied as a selection cut for background reduction and additional background measurements at KKL will further reduce the background and its model uncertainty, the largest systematic on $N_s$. Neutron irradiation campaigns for the energy calibration at threshold, quenching-factor measurements in germanium with improved precision at sub-keV nuclear recoil energies, and dedicated active-mass measurements will reduce the dominant uncertainties on the signal prediction~\cite{CONUS:2026run2}. Constraints on physics beyond the SM previously derived from run 1, such as on neutrino electromagnetic properties, non-standard interactions, and light mediators~\cite{CONUS:2026uhz}, will substantially benefit from the run 2 precision, marking the transition of reactor \CEvNS~from first detection towards precision measurements.


\begin{acknowledgments}
\textit{Acknowledgments.}---We thank all the technical and administrative staff who helped in building the experiment, in particular the MPIK workshops and Mirion Technologies (Canberra) SAS in Lingolsheim, France. Our deepest gratitude goes to Kernkraftwerk Leibstadt AG for hosting and supporting the \conusplus~experiment, with special thanks to P.~Oeschger, P.~Spahr, and P.~Marsanik. We thank A.~Onillon, T.~Lasserre and C.~Wiesinger for their careful reading of the manuscript and valuable comments. All authors are members of the CONUS Collaboration.
\end{acknowledgments}

\bibliography{references}{}

\end{document}